%% file: pastur.tex
\documentclass{amsproc1}
\usepackage[final]{epsfig}
\usepackage{bbm,amssymb,amsmath,amsthm,amsbsy,amsfonts,times}
\usepackage{submitextras}
\numberwithin{equation}{section}
\newtheorem{theorem}{Theorem}[section]

\theoremstyle{definition}
\newtheorem{remark}[theorem]{Remark}
\newtheorem{remarks}[theorem]{Remarks}
\newtheorem{definition}[theorem]{Definition}
\newtheorem{conjecture}[theorem]{Conjecture}
\begin{document}
\title[Random Schr\"odinger operators for amorphous solids]{A survey of rigorous results on random Schr\"odinger operators for amorphous solids}

\author{Hajo Leschke} %
             \address{Institut f\"ur Theoretische Physik, 
                       Universit\"at  
                       Erlangen-N\"urnberg, Staudtstra{\ss}e 7, D--91058 
                       Erlangen, Germany}
             \email{hajo.leschke@physik.uni-erlangen.de}%
\author{Peter M\"uller}
\address{Institut f\"ur Theoretische Physik, 
                       Georg-August-Universit\"at G\"ottingen, 
                       Tammannstra{\ss}e 1, D--37077 G\"ottingen, Germany} \email{peter.mueller@physik.uni-goettingen.de}%
\author{Simone Warzel} %
\address{Institut f\"ur Theoretische Physik, 
                       Universit\"at  
                       Erlangen-N\"urnberg, Staudtstra{\ss}e 7, D--91058 
                       Erlangen, Germany}%
\email{simone.warzel@physik.uni-erlangen.de}%

\begin{abstract}
  Electronic properties of amorphous or non-crystalline disordered solids are often modelled by one-particle Schr\"odinger operators
  with random potentials which are ergodic with respect to the full group of Euclidean translations. 
  We give a short, reasonably self-contained survey of rigorous results on such operators, where we allow for the presence of a 
  constant magnetic field.  
  We compile robust properties of the integrated density of states like its self-averaging, uniqueness
  and leading high-energy growth. 
  Results on its leading low-energy fall-off, that is, on its Lifshits tail, are then discussed 
  in case of Gaussian and non-negative Poissonian random potentials.  
  In the Gaussian case with a continuous and non-negative covariance function 
  we point out that the integrated density of states is locally Lipschitz continuous 
  and present explicit upper bounds on its derivative, 
  the density of states. 
  Available results on Anderson localization concern the almost-sure pure-point nature of 
  the low-energy spectrum 
  in case of certain Gaussian random potentials for arbitrary space dimension. 
  Moreover, under slightly stronger conditions all absolute spatial moments of an initially localized wave packet in the pure-point 
  spectral subspace 
  remain
  almost surely finite for all times.
  In case of one dimension and a Poissonian random potential with repulsive impurities of finite range,
  it is known that the whole energy spectrum is almost surely only pure point.
\end{abstract}

\maketitle

\begin{center}
Dedicated to Leonid A.\ Pastur on the occasion of his $ 65 $th birthday. 
\end{center}

%
\setcounter{tocdepth}{2}
\tableofcontents

\section{Introduction}
Over the last three decades a considerable amount of rigorous results on random Schr\"odinger operators 
have been achieved by many researchers.
Good general overviews of such results 
can be found in the review articles \cite{Spe86,MaSc87,Kir89,Pas99,Sto02,Ves03}
and the monographs \cite{CyFr87,CaLa90,PaFi92,Sto01}. 
We also recommend these sources for the general background in the field.

Most works concern Schr\"odinger operators with random potentials that possess  
an underlying lattice structure even if they are defined on continuous space. 
As against that, the present survey
aims to collect rigorous results on one-particle Schr\"odinger operators 
with (and only with) \emph{truly continuum} random potentials
modelling \emph{amorphous solids}.
For the sake of simplicity we will refrain from stating these results 
under the weakest assumptions available for their validity, but only provide 
sufficient conditions which are easy to check. 
For weaker assumptions and related slightly stronger results the interested reader is
referred to the original cited works where also the corresponding proofs can be found, which will be omitted here. 
As far as results by the present authors are mentioned, they have been obtained 
in collaboration with Jean-Marie Barbaroux, Markus B{\"o}hm, Kurt Broderix~(1962--2000), Werner Fischer, Nils Heldt, 
Dirk Hundertmark, Thomas Hupfer and/or Werner Kirsch.\\ 

It is a pleasure to dedicate this survey to Leonid A.\ Pastur -- one of the founding fathers of the rigorous theory
of disordered systems. He is a mathematical physicist who masterly knows how to convert physical
intuition into mathematical theorems and vice versa. 
Many of his contributions to the theory of random Schr\"odinger operators have been ground breaking.
Here we only mention the early papers \cite{BePa70,Pas71,Pas72,Pas77,GoMo77,Pas80}, his survey articles \cite{Pas73,Pas99} 
and the monographs \cite{LiGr88,PaFi92}.
\subsection{Motivation and models} 
Almost half a century after Anderson's pioneering paper~\cite{And58},
one-particle Schr\"odinger operators with random potentials
continue to play a prominent r\^ole for understanding electronic properties of disordered solids.
While perfect solids or crystals are characterized by the periodic arrangement of identical atoms (or ions) on the sites
of a lattice, disordered solids lack any kind of long-range order, but may 
exhibit some vestige of short-range order.
In some disordered solids, like in random alloys, an underlying lattice structure may still exist so that
on average the solid remains homogeneous with respect to lattice translations.
Accordingly, the random potentials employed to model such solids should be lattice-homogeneous, or even lattice-ergodic,
in order to  also take into account that different parts of the solid are practically decoupled at large separation \cite{BoEn84,LiGr88}.
In extremely disordered materials, like liquids, glasses or amorphous solids~\cite{Zal83,Kit96,GeSm01,KKK01}, 
an underlying lattice structure is no longer available and the corresponding random potentials 
should even be ergodic with respect to the group of \emph{all} Euclidean translations,
not only lattice ones. 

The present survey is concerned with one-particle Schr\"odinger operators with random potentials modelling amorphous solids. 
More generally, we will consider a single quantum particle in $ d $-dimensional Euclidean 
configuration space $ \mathbbm{R}^d $, $ d \in \mathbbm{N} $, subject to a random potential in the sense of 
\begin{definition}\label{Def:pot}
By a \emph{random potential} we mean a 
random scalar field $ V : \Omega \times \mathbbm{R}^d \to \mathbbm{R}$, 
$\,  x \mapsto V^{(\omega)}(x) $ 
on a complete probability space $ \left( \Omega, \mathcal{A}, \mathbbm{P} \right) $, 
such that $ V $ is jointly measurable with respect to 
the product of the sigma-algebra $ \mathcal{A} $ of (event) subsets of $\Omega$ 
and the sigma-algebra $\mathcal{B}(\mathbbm{R}^d)$ 
of Borel sets in $ \mathbbm{R}^d $.  
We further suppose that $ V $ is $ \mathbbm{R}^d $-ergodic and that it has one of the following two
properties:
\begin{indentnummer*}
\item[\ass{V--}] $ V $ has a finite $ p $th absolute moment, that is, 
  $ \mathbbm{E}\big[ \left| V(0) \right|^p \big] < \infty $ 
  with some real $ p > \max\left\{ 3 , d +1\right\} $.
\item[\ass{V+}] $ V $ is non-negative and has a finite $ p $th moment, that is,
  $ \mathbbm{E}\big[ V(0)^p \big] < \infty $
  with $ p = 2 $ if $ d \in \{ 1, 2 , 3 \} $ or with some real $ p > d/ 2 $ if $ d \geq 4 $.
\end{indentnummer*}
Here $ \mathbbm{E}[ \cdot ] := \int_\Omega\mathbbm{ P}( \d \omega) \, (\cdot) $ denotes the \emph{expectation} 
(in other words: ensemble averaging) 
induced by the probability measure $ \mathbbm{ P} $.
\end{definition}

The precise definition of \emph{$ \mathbbm{R}^d $-ergodicity} of $ V $ 
requires \cite{Kre85,Kir89} the existence of a group $ \{ \mathcal{T}_x \}_{x \in \mathbbm{R}^d} $
of probability-preserving transformations on $ \left( \Omega, \mathcal{A}, \mathbbm{P} \right) $ such that 
~(i)~ the group is ergodic in the sense that 
every event $ \mathcal{E} \in \mathcal{A} $ which is invariant under the whole group is 
either almost impossible or almost sure, $ \mathbbm{P}(\mathcal{E}) \in \{0,1\} $,
and that ~(ii)~ $ V $ is \emph{$ \mathbbm{R}^d $-homogeneous} in the sense that 
$ V^{(\mathcal{T}_x \omega)}(y) =  V^{(\omega)}(y- x) $ for (Lebesgue-) almost 
all $ x $, $ y \in \mathbbm{R}^d $ and all $ \omega \in \Omega $.
Roughly speaking, $ V $ is $ \mathbbm{R}^d $-ergodic if its fluctuations in different regions become sufficiently fast 
decorrelated with increasing distance between the regions.\\

The quantum particle subject to 
the random potential may also be exposed to a \emph{constant magnetic field} 
characterized by a skew-symmetric $ d \times d $-matrix 
with real entries $ B_{ij} = - B_{ji} $, where $ i, j \in \{1, \dots , d\} $. The components of the corresponding vector potential 
$ A : \mathbbm{R}^d \to \mathbbm{R}^d $ in the Poincar\'e gauge are 
defined by $ A_j(x) := \frac{1}{2} \sum_{i=1}^d x_i B_{ij} $ for all $ x = (x_1, \dots , x_d) \in \mathbbm{R}^d $.

Choosing physical units such that the mass and electric charge of the particle
as well as Planck's constant (divided by $ 2 \pi $) are all equal to one, the Schr\"odinger operator for the quantum 
particle subject to 
a \emph{realization} $ V^{(\omega)} : \mathbbm{R}^d \to \mathbbm{R} $, 
$ x \mapsto V^{(\omega)}(x) $ of a random potential $ V $ and a constant magnetic field is informally given by the 
differential expression
\begin{equation}\label{re:schop}
  H\big(A,V^{(\omega)}\big) := \frac{1}{2} \sum_{j=1}^d \Big( \i \, \frac{\partial \;}{\partial x_j} +  A_j \Big)^2 + V^{(\omega)},
\end{equation}
where $ \i = \sqrt{-1} $ is the imaginary unit. 
According to the basic postulates of quantum mechanics, this expression has to be well defined as a self-adjoint operator acting on a dense domain 
in $ \mathrm{L}^2(\mathbbm{R}^d) $, the Hilbert space of all complex-valued, Lebesgue square-integrable functions on $ \mathbbm{R}^d $,
which is equipped with the usual scalar product 
$ \langle \varphi  ,  \psi \rangle := \int_{\mathbbm{R}^d} \d x \, \varphi(x)^* \psi(x) $, 
for $ \varphi$, $ \psi \in \mathrm{L}^2(\mathbbm{R}^d)$.
In fact, our assumptions on $ V $ guarantee the existence of some subset $ \Omega_0 \in \mathcal{A} $ of $ \Omega $ 
with full probability, $ \mathbbm{P}(\Omega_0) = 1 $, such that for every $ \omega \in \Omega_0 $ the right-hand side of (\ref{re:schop})
is essentially self-adjoint on the dense subspace 
$ \mathcal{C}_0^\infty(\mathbbm{R}^d) \subset  \mathrm{L}^2(\mathbbm{R}^d) $ of all complex-valued, arbitrarily often differentiable
functions with compact supports, 
see for example \cite{KirMar83a,Kir89,CaLa90,FiLeMu00}. 
This justifies
\begin{definition}
By a \emph{random Schr\"odinger operator} $ H(A,V) $ with a random potential $ V $ and 
a constant magnetic field, corresponding to the vector potential $ A $, 
we mean the family $ \Omega_0 \ni \omega \mapsto H(A,V^{(\omega)}) $
of Schr\"odinger operators given by~(\ref{re:schop}).\\
\end{definition}

In this survey we basically focus on two examples of random potentials in the sense of Definition~\ref{Def:pot}, namely Gaussian and non-negative Poissonian ones. 
Both are rather popular in the physics literature \cite{Zi79,BoEn84,ShEf84,LiGr88,Jan94,Efe97}, see also \cite{KM93}
and references therein.  
\begin{definition}
  By a \emph{Gaussian random potential} we mean a Gaussian random field~\cite{Adl81,Lif95}
  which is $ \mathbbm{R}^d $-ergodic.
  It has  zero mean, $ \mathbbm{E}\left[\,V(0)\right] = 0 $, 
  and its covariance function
    $ C: \mathbbm{R}^d \to \mathbbm{R} $, $ x \mapsto C(x):= \mathbbm{E}\left[\, V(x) V(0) \right] $ 
    is continuous at the origin where it obeys 
    $ 0 < C(0) < \infty $.
  \end{definition}
The covariance function $ C $ of a Gaussian random potential 
is bounded and 
uniformly continuous on $ \mathbbm{R}^d $ by definition. 
Consequently, \cite[Thm.\ 3.2.2]{Fer75} implies the existence of 
a separable version of this field which is jointly  measurable.
Referring to a Gaussian random potential,
we will tacitly assume that only this version will be dealt with.
By the Bochner-Khintchine theorem 
there is a one-to-one correspondence
between Gaussian random potentials and finite positive (and even)
Borel measures on $ \mathbbm{R}^d $. 
A simple sufficient criterion ensuring 
$ \mathbbm{R}^d $-ergodicity
is the mixing condition $\lim_{|x| \to \infty} C(x) = 0$. 
Furthermore, the explicit formula 
\begin{equation}\label{eq:explicit}
  \mathbbm{E}\big[ \left| V(0) \right|^p \big] 
  =  (2 \pi C(0))^{-1/2} 
  \int_{\mathbbm{R}} \!\! \d v \; \e^{-v^2/2 C(0)} \, | v |^p \, 
  = \Gamma\left(\frac{p+1}{2}\right) \, \frac{\left[ 2 C(0) \right]^{p/2}}{\pi^{1/2}},  
\end{equation}
where $ \Gamma $ stands for 
Euler's gamma function \cite{GrRy}, shows that a Gaussian random potential has property~\ass{V--} and is therefore
a random potential in the sense of Definition~\ref{Def:pot}.\\

The second example of a random potential considered subsequently is a non-negative Poissonian one, 
which we define as follows.
\begin{definition}
  By a non-negative \emph{Poissonian random potential} with \emph{single-impurity potential} 
  $ U \geq 0 $ and \emph{mean concentration} $ \varrho > 0 $ 
  we mean a random field with realizations given by
  \begin{equation}\label{def:pois}
    V^{(\omega)}(x) = \int_{\mathbbm{R}^d} \! \mu_{\varrho}^{(\omega)}(\d y) \; U(x-y). 
  \end{equation}
  Here $ \mu_\varrho $ denotes the (random) Poissonian measure on $ \mathbbm{R}^d $ 
  with parameter $ \varrho > 0 $ and the function 
  $ U: \mathbbm{R}^d \to [0,\infty[ $ is supposed to be
  non-negative and strictly positive on some non-empty open set in $ \mathbbm{R}^d $. 
  Moreover, we assume that $U $ 
  satisfies the Birman-Solomyak condition 
  $ \sum_{m \in \mathbbm{Z}^d} \big( \int_{\Lambda_m} \d x \; |U(x)|^p \big)^{1/p} < \infty $ 
  with  $ p = 2 $ if $ d \in \{ 1, 2 , 3 \} $ or with some real $ p > d/ 2 $ if $ d \geq 4 $.
  Here $ \Lambda_m $ denotes the unit cell of the $ d $-dimensional simple cubic \emph{lattice}
  $ \mathbbm{Z}^d \subset \mathbbm{R}^d $ 
  which is centred at the site $ m \in \mathbbm{Z}^d $.
\end{definition}

Since the Poissonian measure is a random Borel measure which is only pure point and positive integer-valued, 
each realization of a non-negative Poissonian
random potential is informally given by 
\begin{equation}
  V^{(\omega)}(x)= \sum_j U\big(x - p_{j}^{(\omega)} \big).
\end{equation} 
It can therefore be interpreted as the potential generated by immobile \emph{impurities}, 
located at $ \{ p_{j}^{(\omega)} \} \subset \mathbbm{R}^d $, each of which is characterized by the same repulsive  
potential $ U $. The random variable $ \mu_{\varrho}(\Lambda) $ then equals the number of 
impurities in the bounded Borel set $ \Lambda \subset \mathbbm{R}^d $. It is Poissonian distributed according to 
\begin{equation} 
  \mathbbm{P}\left(\big\{ \, \omega \in \Omega \, \big| \,  \mu_{\varrho}^{(\omega)}(\Lambda) = n \, \big\}\right)
  = \frac{\left( \varrho \left| \Lambda \right| \right)^{n}}{n!} \, 
  \e^{ - \varrho \left| \Lambda \right|}, \qquad  n \in \mathbbm{N} \cup \{ 0 \},
\end{equation}
where $ | \Lambda | := \int_\Lambda \d x $ denotes the (Lebesgue-) volume of $ \Lambda $, so that
$ \varrho $ is indeed the mean (spatial) concentration of impurities.
Employing for example \cite[Lemma~3.10]{HuLeMuWa01b}, one makes sure that a non-negative Poissonian random potential
satisfies property~\ass{V+}. Since it is also $ \mathbbm{R}^d $-ergodic due to $ U \in \mathrm{L}^2(\mathbbm{R}^d) $, it is therefore a random potential 
in the sense of Definition~\ref{Def:pot}.

\subsection{Interesting quantities and basic questions} 
A quantity of primary interest in the theory and applications 
of random Schr{\"o}dinger operators is the (specific) integrated density of states $ N $. 
Its knowledge allows one to compute the specific free energy of the corresponding non-interacting 
many-particle system in the thermodynamic limit, see for example (\ref{eq:free}) below. 
It also enters formulae for transport coefficients.

For its definition we first introduce $ \Theta( E - H(A,V^{(\omega)})) $, the spectral projection operator of $ H(A,V^{(\omega)}) $
associated with the open half-line  $ ]-\infty , E \, [ \, \subset \mathbbm{R} $ up to a given energy $ E \in \mathbbm{R} $.
This notation complies with the functional calculus for self-adjoint operators in that  
$ \Theta: \mathbbm{R} \to \{0,1\}\, $ stands for Heaviside's 
left-continuous unit-step function. 
For a random potential in the sense of Definition~\ref{Def:pot}, this spectral projection possesses a 
complex-valued integral kernel (in other words: position representation)
$ \Theta( E - H(A,V^{(\omega)}))(x,y) $, which is a jointly continuous function of $ x $, $ y \in \mathbbm{R}^d $
and a $ \mathbbm{P} $-integrable function of $ \omega \in \Omega_0 $, see~\cite{Sim82,BrHuLe00,BrMuLe01}. 
This justifies
\begin{definition}
The \emph{integrated density of states} is the function $ N: E \mapsto N(E) $ defined through the expectation value 
\begin{equation}\label{Def:N}
  N(E) := \mathbbm{E} \left[ \Theta\big( E - H(A,V)\big)(x,x) \right]  
  = \! \int_{\Omega} \! \mathbbm{P}( \d \omega) \,\; \Theta\left( E - H(A,V^{(\omega)})\right)(x,x).
\end{equation} 
$\mbox{}$
\end{definition}
Thanks to the unitary invariance of the kinetic-energy operator $ H(A,0) $ under so-called magnetic translations~\cite{Zak64}
and to the $\mathbbm{R}^d $-homogeneity of $ V $, $ N $ 
is independent of the chosen~$ x \in \mathbbm{R}^d $. 
Moreover, $ N $ is non-negative, non-decreasing and left-continuous.

There are some other universally valid properties of 
the integrated density of states $ N $ which do not depend on the 
specific choice of the random potential. 
For example, for theoretical and experimental reasons it is a comforting fact to learn that $ N $, 
which is defined above as an ensemble average involving the infinite-volume random operator $ H(A,V) $,
may be viewed as a spatial average for a given typical realization $ V^{(\omega)} $ of $ V $. 
This property, often dubbed \emph{self-averaging} \cite{LiGr88}, and the arising \emph{uniqueness} problem 
will be made more precise in Section~\ref{Sec:IDOS} below.
Basically, self-averaging is a consequence of the assumed ergodicity of the random potential.
The latter is also responsible for the almost-sure non-randomness of the spectrum of $ H(A,V) $
and of its spectral components in the Lebesgue decomposition~\cite{Pas80,KirMar82a}. 
By~(\ref{Def:N}), the location of the almost-sure spectrum of $ H(A,V) $, as a closed subset of the real line $ \mathbbm{R} $, 
coincides with the 
\emph{set of growth points} $ \{ E \in \mathbbm{R}  \, | \,  N(E) < N(E + \varepsilon ) \; \mbox{for all} \; \varepsilon > 0 \} $ of $ N $.
But the location of the spectral components (in other words: the nature of the spectrum) 
cannot be inferred from $ N $ alone. For rather simple exceptions see the paragraph below (\ref{Eq:Theophys}). 
In any case, the spectrum depends on the choice of $ V $. 
For example, due to the unboundedness of the negative fluctuations of a Gaussian random potential $ V $, 
the almost-sure spectrum of the corresponding $ H(A,V) $ coincides with the whole real line.
In case of a non-negative Poissonian potential the spectrum of   
$ H(A,V) $ is almost surely equal to the half-line starting at the ground-state energy of $ H(A,0) $, that is, 
at the infimum of its spectrum.

Another universally valid property of $ N $ is its leading high-energy growth which is given by
\begin{flalign}\label{Eq:high}
  &&
  N(E) \sim \frac{1}{\Gamma(1+d/2)} \left(\frac{E}{2\pi}\right)^{d/2} 
  &&& (E \to \infty).
\end{flalign}
see \cite{Kir89,CaLa90,PaFi92,Mat93,Uek94}. 
Here we use the notation 
$ f(E) \sim g(E) $ as $ E \to E' $ 
to indicate the asymptotic equivalence in the sense that $ \lim_{E\to E'} f(E)/ g(E) = 1 $ for some $ E' \in [-\infty,\infty] $. 
Remarkably, the asymptotics (\ref{Eq:high}) is 
purely classical in the sense that
$ N(E) \sim N_c(E) $ as $ E  \to \infty $, where  
\begin{align}\label{DefclIDOS}
  N_c(E)  := & \, \frac{1}{\left|\Lambda\right|} \, \mathbbm{E} \left[
    \int_{\Lambda \times \mathbbm{R}^d} \! \frac{\d x \, \d k}{(2 \pi)^d} \;\;
    \Theta\Big(E - \tfrac{1}{2}\left| k - A(x) \right|^2 - V(x) \Big) 
  \right] \\
  = & \, \frac{1}{\Gamma(1+d/2)} \, 
  \mathbbm{E} \left[\left( \frac{E-V(0)}{2 \pi}\right)^{d/2} \, 
    \Theta\big(E- V(0)\big)\right] \notag 
\end{align}
defines the \emph{(quasi-) classical integrated density of states}, see also \cite{Kan83}.
In accordance with a theorem of Bohr and van Leeuwen
on the non-existence of diamagnetism in classical physics \cite{Mat88},
the integration with respect to the canonical momentum $k \in \mathbbm{R}^d$ 
shows that $N_c $ does not depend on $ A $ and hence not on the magnetic field $ (B_{ij}) $.
Furthermore, the $ \mathbbm{R}^d $-homogeneity of $ V $ ensures that $ N_c $ is independent of the chosen 
bounded Borel set $ \Lambda \subset \mathbbm{R}^d $ of strictly positive volume $ | \Lambda | $.
The asymptotics (\ref{Eq:high}) does not even depend on the random potential. Rather it is 
consistent with
a famous result of Weyl~\cite{Wey12}. 

In contrast to the high-energy growth (\ref{Eq:high}), the low-energy fall-off of $ N $ near the 
almost-sure ground-state energy of $ H(A, V) $ is not universal, more complicated and in general harder to obtain.
It typically stems 
from exponentially rare low-energy fluctuations of the random potential, 
which compete with the quantum fluctuations related to the kinetic energy.
As a result, $ N $ exhibits a much faster low-energy fall-off in comparison to the
non-random case $ V = 0 $. This is commonly referred to as a \emph{Lifshits tail} in honour
of the theoretical physicist I.\ M.\ Lifshits (1917--1982) who was the first to develop a quantitative theory~\cite{Lif63,Lif64}
in case $ A = 0 $.
Lifshits' arguments can be summarized in terms of the so-called
\emph{optimal-fluctuation ideology}~\cite{HaLa66,ZiLa66,Lif67} 
(see also \cite{ShEf84,LiGr88,PaFi92}) according to which 
the low-energy fall-off of $ N $ near the almost-sure
ground-state energy $ E_0 \in [-\infty, \infty [$ of $ H(A, V) $ is (universally) given by the formula
\begin{flalign}\label{Eq:Theophys}
  &&
  \quad \log N(E) 
  \sim  \inf_{\tau >0} \Big( \, \tau E + \!\!\!
  \sup_{\substack{\psi \in  \mathcal{C}_0^\infty(\mathbbm{R}^d) \\
      {\langle\psi,\, \psi\rangle= 1}}}
  \!\!\! \log \, \mathbbm{E} \big[  \e^{-\tau \langle \psi \, , \, H(A,V) \psi \rangle} \big]\Big)
  && (E \downarrow E_0).
\end{flalign}
Here and in the following we suppress suitable constants ensuring dimensionless arguments of the natural logarithm 
because they become irrelevant in the limit.
Actually all Lifshits tails we know of, in particular the ones caused by Gaussian and Poissonian random
potentials presented in Sections~\ref{Sec:Gauss} and~\ref{Sec:Poisson} below, are consistent with (\ref{Eq:Theophys}).

As a left-continuous (unbounded) distribution function, $ N $ has at most countably many discontinuity points. 
For one space dimension, $ N $ is known to be even globally continuous \cite{Pas80}. 
If this is also true in the multi-dimensional case (of a continuum and $ A = 0 $), is an open problem.
Using (\ref{Def:N}) it is not hard to show (for example by \cite[Cor.~3.7]{HuLeMuWa01b} and \cite[Thm.~2.12]{PaFi92}) 
that $ N $ is discontinuous at a given energy $ E \in \mathbbm{R} $ if and only if $ E $ 
is almost surely an infinitely degenerate eigenvalue of $ H(A,V) $.
In the other extreme case in which $ H(A,V) $ has almost surely only absolutely continuous spectrum 
in a given bounded energy interval $ I \subset \mathbbm{R} $, it follows directly from (\ref{Def:N}) 
that $ N $ is absolutely continuous on $ I $. This means 
that the \emph{density of states} $ D: E \mapsto D(E) := \d N(E) / \d E $ 
exists as a non-negative Lebesgue-integrable function on $ I $. 
In any case, under more specific assumptions on $ V $ regularity of $ N $ 
beyond absolute continuity is expected on the whole almost-sure energy spectrum, 
even in regimes with (dense) pure-point spectrum.
For example, in case of a Gaussian random potential with a non-negative covariance function, 
$ N $ turns out to be (at least) locally Lipschitz
continuous, so that $ D $ exists on the whole real line and is locally bounded, see Theorem~\ref{Thm:DOS} below. 
The regularity of $ N $ is physically relevant, for instance, 
for the basic thermal-equilibrium properties of the corresponding macroscopic system of 
non-interacting (spinless) fermions in a random medium. These properties are determined by the \emph{specific free energy}
\begin{equation}\label{eq:free}
  F(T, \bar{n}) := \sup_{\mu \in \mathbbm{R}} \left[ \, \mu \, \bar{n} 
    - T \int_{\mathbbm{R}} \!\! \d N(E) \, \log \left( 1 + \e^{(\mu - E)/T} \right) \right]
\end{equation}
as a function of the absolute temperature $ T > 0 $ (multiplied by Boltzmann's constant), the spatial concentration 
$ \bar{n} > 0 $ of the fermions and (possibly) the magnetic field. 
In fact, Sommerfeld's ubiquitous asymptotic low-temperature expansion of $ F $ presupposes 
a sufficiently smooth integrated density of states \cite{Som28}, \cite[Chap.~4]{Kub65}. 
In particular, one then has  
$ F(T, \bar{n}) \sim  $\hspace{0pt}$\int_{- \infty}^{E_F} \d E \, D(E) E  -  T^2 D(E_F) \, \pi^2 / 6 $ as $ T \downarrow 0 $,  
where the \emph{Fermi energy} $ E_F \in \mathbbm{R} $ is the solution of the equation $ N(E_F) = \bar{n} $. 
Given (\ref{Eq:high}), a sufficient condition for the existence of the Lebesgue-Stieltjes integral in (\ref{eq:free}) is the finiteness 
$ \mathbbm{E}\left[ \exp\left( - \tau V(0) \right) \right] < \infty $ for all $ \tau \in [0, \infty[ $.
In fact, this condition implies the (quasi-) classical estimate $ N(E) \leq \left(2 \pi \tau \right)^{-d/2} e^{\tau E} \, \mathbbm{E}\left[ \exp\left( - \tau V(0) \right) \right] $ 
for all $ E \in \mathbbm{R} $ with arbitrary $ \tau \in ] 0 , \infty [ $ by \cite[Thm.~9.1]{PaFi92} and the diamagnetic inequality, 
see also \cite{BrMuLe01}. \\

In perfect solids or crystals the (generalized) one-electron energy eigenfunctions
are given by Bloch-Floquet functions on $ \mathbbm{R}^d $, which are (time-independent) plane waves modulated by
lattice-periodic functions \cite{Kit96,Kuc93}.
Accordingly these eigenfunctions are delocalized over the whole solid, hence not square-integrable, and the
whole energy spectrum is only absolutely continuous.
That the spectrum is absolutely continuous even in the low-energy regime, may be viewed as a consequence of the tunnelling effect. 
According to classical mechanics the electron would be localized in one of the identical atomic potential wells
making up the crystal.
Since even small differences in the potential wells may suppress ``quantum coherence'' and hence tunnelling, 
localized states given by square-integrable energy eigenfunctions associated with pure-point spectrum should emerge 
in disordered solids at least at low energies.
In particular, this should be true for amorphous solids. 
As we will see in Theorem~\ref{Thm:loc} and Theorem~\ref{thm:ueki} below, localization at low energies 
can indeed be proven in case of certain Gaussian random potentials for arbitrary $ d \in \mathbbm{N} $.

Since by their very nature, localized states are not capable of contributing to macroscopic charge transport,
at a certain (non-random) energy a \emph{mobility edge} 
is expected to occur \cite{Mot67} 
in an amorphous solid which separates localized states in the Lifshits-tail
regime from delocalized ones at higher energies.
At zero temperature, the passing of the Fermi energy through a mobility edge from delocalized to
localized states results in a metal-insulator
transition known as the Anderson transition. 
Likewise, the phenomenon of localization by disorder is
called \emph{Anderson localization} \cite{And58}, see also \cite{KM93} and references therein.

In order to better understand the suppression of charge transport
in disordered solids by Anderson localization,
both spectral and dynamical criteria for localization are commonly
studied. \emph{Spectral localization} means that there is only (dense) pure-point
spectrum in certain energy regimes. In addition, the corresponding
eigenfunctions are often required to decay not slower than exponentially at
infinity instead of being merely square-integrable. 
One criterion for \emph{dynamical localization} is
a sufficiently slow spatial spreading of initially localized 
wave functions 
which evolve under the unitary time evolution in the spectral subspace $ \indfkt{I}( H(A, V^{(\omega)})) \, {\rm L}^2(\mathbbm{R}^d) $ 
corresponding to a certain (Borel) 
energy regime $ I \subset \mathbbm{R} $. 
More precisely, one requires the finiteness 
\begin{equation}\label{eq:dyn}
  \sup_{t \in \mathbbm{R}} \, \int_{\mathbbm{R}^d} \!\! \d x \; \big| \psi_{t,I}^{(\omega)}(x) \big|^2 \, | x |^q \, < \infty 
\end{equation}
of the $ q $th absolute spatial moment of $ \psi_{t,I}^{(\omega)} := \e^{-\i t H(A, V^{(\omega)})} \indfkt{I}( H(A, V^{(\omega)}) ) \psi_0 $ 
for all times $ t \in \mathbbm{R} $, all $ \psi_0 \in \mathcal{C}_0^\infty(\mathbbm{R}^d) $ and, for example, 
a real $ q \geq 2 $. Usually (\ref{eq:dyn}) is demanded either for 
$ \mathbbm{P} $-almost all $ \omega \in \Omega $ or even upon integration over $ \omega $ 
with respect to $ \mathbbm{P} $, see \cite{GeDBi98,BaFi99,DaSt01,GeKl01} and \cite[Sec.~3.4]{Sto01}.
While this type of dynamical localization implies spectral localization in $ I $ by the RAGE theorem \cite[Sec.~5.1]{Amr81} 
(see also \cite{CyFr87}, \cite[Sec.~4.1.5]{Sto01}, \cite{Wei03}), 
the converse is not generally true, since pure-point spectrum may occur together  
with a sub-ballistic long-time behaviour \cite{DRJi95}. 
Even more physical is another criterion for dynamical localization, namely the vanishing of the direct-current
conductivity of the corresponding non-interacting fermion system at zero temperature. 
The interrelations between the different localization criteria 
are complicated and not yet understood in sufficient generality; in this context we recommend \cite{Sim90,DRJi96,Las96,BaCo97c,KiLa99,Tch01}.
\subsection{Random Landau Hamiltonian and its single-band approximation}
In recent decades the physics of (quasi-) two-dimensional 
electronic structures has attracted considerable attention \cite{AnFoSt82,vKl86,KuMeTi88,StTsGo99}.
Some of the occurring phenomena, like the integer quantum Hall effect,
are believed to be microscopically explainable in terms of a 
system of non-interacting electrically charged fermions in the Euclidean plane $ \mathbbm{R}^2 $ subject to a
perpendicular constant magnetic field of strength $ B := B_{12} > 0 $
and a random potential~\cite{Jan94,Yos02}. 
The underlying one-particle Schr\"odinger operator is known as the \emph{random Landau Hamiltonian} which acts on the
Hilbert space $ {\rm L}^2(\mathbbm{R}^2) $.
Apart from numerous theoretical and numerical studies in the physics literature, there are nowadays
quite a lot of rigorous results available for this and related models, 
see for example \cite{Kun87,MaPu92,BeElSc94,CoHi96,BaCoHi97b,Wan97,AiGr98,DoMa99,AvSa01,ElGr02,GeKl02} and references therein.

The kinetic-energy part of the random Landau Hamiltonian 
is the well understood \emph{Landau Hamiltonian}
\begin{equation}\label{eq:Landau}
 H(A,0) = \frac{1}{2} \left[ %
                         \left(\i \frac{\partial}{\partial x_1} -%
                             \frac{B}{2}x_2\right)^2 +%
                         \left(\i \frac{\partial}{\partial x_2} +%
                             \frac{B}{2}x_1\right)^2 %
                       \right] 
        = B \, \sum_{l=0}^\infty \left( l + \frac{1}{2} \right) P_l.
\end{equation} 
Its spectral resolution expressed by the second equality dates back to Fock~\cite{Foc28} 
and Landau~\cite{Lan30}. The energy eigenvalue $(l + 1/2) B$ is called 
the $l$th \emph{Landau level} and the corresponding orthogonal eigenprojection $P_l$ is an
integral operator with continuous complex-valued kernel
\begin{equation}
P_l(x,y) := \frac{B}{2\pi}
  \exp\left[ \i \frac{B}{2}  (x_2 y_1 - x_1 y_2) - \frac{B}{4} |x-y|^2\right] 
  {\rm L}_l\left(\frac{B}{2} |x-y|^2\right)
\end{equation}
given in terms of the $ l $th Laguerre polynomial 
$ {\rm L}_l(\xi):= \frac{1}{l!} \e^\xi \frac{d^l}{d\xi^l} \big( \xi^l \e^{-\xi} \big) $, $\xi \geq 0 $ \cite{GrRy}.
The diagonal $P_l(x,x)= B/2\pi$ is naturally interpreted as the specific degeneracy of the $ l$th Landau level.

Accordingly, the corresponding integrated density of states is the well-known 
discontinuous staircase function
\begin{equation}
  N(E) = \frac{B}{2 \pi} \sum_{l=0}^\infty \Theta\big( E - \left( l + 1/2 \right) B \big), \qquad V = 0.
\end{equation}
Informally, the associated density of states $ \d N(E) / \d E $ is a series of Dirac delta functions supported at the Landau levels.
By adding a random potential to (\ref{eq:Landau}) the corresponding peaks are expected to be smeared out.
In fact, in case of Gaussian random potentials with a non-negative covariance function they are smeared out completely, 
see Theorem~\ref{Thm:DOS} in case $ d = 2 $ and Figure~\ref{bild} below.\\

In the limit of a strong magnetic field the spacing 
$ B $
of successive Landau levels approaches infinity and the so-called \emph{magnetic length}
$ 1/ \sqrt{B }$ tends to zero.
Therefore the effect of so-called level mixing should be 
negligible if either the strength of the
random potential $ V $ is small compared to the level 
spacing or if the (smallest) correlation length
of $ V $ is much larger than the magnetic length.
In both cases a reasonable approximation to $ N $
should then read 
\begin{equation}\label{eq:decomp}
  N(E) \approx \frac{B}{2\pi} \sum_{l = 0}^{\infty}  R_l\big( E - \left( l + 1/2 \right) B \big).
\end{equation}
Here the $ l $th term of the infinite series stems from the probability distribution function $ R_{l} $ on $ \mathbbm{R} $
defined by
\begin{equation}\label{Def:Rl}
  R_{l}(E) := \frac{2 \pi}{B} \, \mathbbm{E} \left[ \big( P_l \, \Theta\left(E - P_l V P_l \right) P_l \big)(x,x) \right],
  \qquad x \in \mathbbm{R}^2.
\end{equation}
We refer to it as the (centred and normalized) $ l $th 
\emph{restricted integrated density of states}. 
It describes the broadening  
of the $ l $th Landau level by the random potential to the $ l $th 
\emph{Landau band}, when considered in isolation. 
Assuming (without loss of generality) that the mean of the $ \mathbbm{R}^2 $-homogeneous random
potential $ V $ is zero, $ \mathbbm{E}\left[ V(0) \right] = 0 $, the variance of the energy 
in the $ l $th Landau band is given by \cite{BrHeLe91}
\begin{equation}\label{Eq:2mom}
        \sigma^2_l := \int_{\mathbbm{R}} \d  R_{l}(E) \, E^2 
        = \frac{2 \pi}{B} \int_{\mathbbm{R}^2} \!\! {\rm d} x \, \left| P_l(0,x) \right|^2 \, C(x)
\end{equation} 
in terms of the covariance function $ C(x) := \mathbbm{E}\left[ V(x) V(0) \right] $ of $ V $.
The standard deviation $ \sigma_l := \sqrt{\sigma_l^2} $ is physically interpreted as the (effective) \emph{width} of the $l$th Landau band.
The exact formula (\ref{Eq:2mom}) first appeared in approximations to $ R_l $, like the so-called
self-consistent Born approximation, see \cite{KuMeTi88} and references therein. 
In case of the \emph{Gaussian covariance function} (with $ x \in \mathbbm{R}^2 $)
\begin{equation}\label{Eq.Gausscov}
  C(x) = C(0) \exp\left(- \frac{|x|^2}{2 \lambda^2} \right), \qquad C(0) > 0, \quad \lambda > 0,
\end{equation}
the $ l $th band width can be calculated exactly and explicitly as a function of the squared length ratio $ B \lambda^2 $.
The results is 
\begin{equation}\label{Eq:levelwidth}
  \sigma^2_l = C(0) \, \frac{B \lambda^2}{B \lambda^2 + 1} \left(\frac{B \lambda^2-1}{B \lambda^2 + 1}\right)^l \, 
  {\rm P}_l\left(\frac{\left(B \lambda^2\right)^2 + 1}{\left(B \lambda^2 \right)^2 - 1}\right) =: w_l\left(B \lambda^2\right) > 0,
\end{equation}
where $ {\rm P}_l(\xi) := \frac{1}{l! 2^l} \frac{d^l}{d \xi^l} \big(\xi^2 - 1\big)^l $, $ \xi \in \mathbbm{R} $, is 
the $ l $th Legendre polynomial \cite{GrRy}.

Neglecting effects of level mixing by only dealing with the sequence of 
restricted random operators $\left\{ P_l V P_l \right\}_l $
is a simplifying approximation which is often made.
The interest in these operators 
relates to spectral localization \cite{Huc95,DoMa95,DoMa96,DoMa97,PuSc97,Scr99} and 
to properties of their (restricted) integrated density of states $ R_l $, see 
for example \cite{Weg83,BeGrIt84,KlPe85,BenCha86,Ben87,Ape87,Sal87,BrHeLe91,BoBrLe97,HuLeWa01}.
%
From the physical point of view most interesting is the restriction 
to the lowest Landau band ($l = 0$).
For strong enough magnetic fields all fermions may be accommodated in the 
lowest band without conflicting with Pauli's exclusion principle,
since the specific degeneracy increases with the magnetic field.
The contribution of $ 2 \pi R_0(E_F - B/2) / B $ to the sum of
the series in (\ref{eq:decomp}) at the Fermi energy $ E_F $ 
should then already be a good approximation to $ N(E_F) $, 
since the effects of higher Landau bands are negligible 
if $ B $ is large compared to the strength $ \sqrt{ C(0) } $ of the
random potential. 
For Gaussian and non-negative Poissonian random potentials, rigorous statements
in support of this heuristics
can be found in \cite{MaPu92,BrHuLe93,War01}.

\begin{remark}
   A sufficient condition \cite{DoMa95,BrHuLe95} for the almost-sure self-adjointness 
   of the restricted random operator $ P_l V P_l $ 
   on $ P_l \mathrm{L}^2(\mathbbm{R}^2) $ is the following
   growth limitation for the even moments of the random potential,
   \begin{equation}\label{eq:mom}
   \mathbbm{E}\left[| V(0) |^{2n} \right] \leq (2 n)! \, M^{2n}
   \end{equation}
   for all $ n \in \mathbbm{N} $ with some constant $ M < \infty $.
   While (\ref{eq:mom}) is satisfied for all Gaussian random potentials with $ M = \sqrt{C(0)} $ (see (\ref{eq:explicit})), 
   its validity for non-negative Poissonian random potentials is ensured \cite{BrHuLe95} by the additional 
   (Lebesgue-essential) boundedness of the 
   single-impurity potential, $ U \in {\rm L}^\infty(\mathbbm{R}^2) $. 
   Moreover, $ R_l $ in (\ref{Def:Rl}) is well defined, because
   the integral kernel $ \big( P_l \Theta(E - P_l V^{(\omega)} P_l) 
   P_l \big)(x,y) $ of the spectral projection $ P_l \Theta\big(E - P_l V^{(\omega)} P_l \big) P_l $
   is jointly continuous in $ x $, $ y \in \mathbbm{R}^2 $ and integrable with respect to $ \mathbbm{P} $ 
   as a function of $ \omega \in \Omega $.  
   Thanks to magnetic translation
   invariance $ R_l $ is independent of the chosen $ x \in \mathbbm{R}^2 $. 
\end{remark}

\section{Self-averaging and uniqueness of the integrated density of states}\label{Sec:IDOS}
Since spatially separated, large parts of a macroscopic sample of an amorphous solid become
decoupled rather fast with increasing distance, they effectively 
correspond to different realizations of the ergodic
random potential modelling the solid. 
As a consequence, it should make no difference
whether the integrated density of states $ N $ is defined as an ensemble average 
or as a spatial average for a given typical realization. 

To specify the notion of a spatial average associated with $ N $, 
one first has to consider a
restriction of the (infinite-volume) random Schr\"odinger operator $ H(A,V) $ 
to a bounded open cube $\Lambda \subset \mathbbm{R}^d$.
The resulting~ ~\emph{finite-volume random Schr\"odinger operator} $ H_{\Lambda,{\rm X}}(A,V) $ 
is rendered almost surely 
self-adjoint
on the Hilbert space $ \mathrm{L}^2(\Lambda) $ by imposing, for example, Dirichlet, $ {\rm X}= {\rm D} $,
or Neumann, $ {\rm X} = {\rm N} $, boundary conditions  
on the wave functions in its domain of definition, see for example \cite{Kir89,CaLa90,HuLeMuWa01a}. 
Since the spectrum of $ H_{\Lambda,{\rm X}}(A,V) $ almost surely consists only
of isolated eigenvalues of finite multiplicity, 
the \emph{finite-volume integrated density of states} $ N_{\Lambda,{\rm X}}^{(\omega)} $ is well defined 
as the (specific) eigenvalue counting-function
\begin{align}
  N_{\Lambda,{\rm X}}^{(\omega)}(E)  := & \frac{1}{|\Lambda |} \, \biggl\{\, 
        \begin{array}{l}
          \text{number of eigenvalues of
            $H_{\Lambda,{\rm X}}(A,V^{(\omega)})$, counting} \\
          \text{multiplicity, which are strictly smaller than $E\in \mathbbm{R}$}
        \end{array}
      \biggr\} \\
      = & \frac{1}{|\Lambda |} \,
      \int_{\Lambda} \! \d x \; \Theta\big( E - H_{\Lambda,{\rm X}}\big(A,V^{(\omega)}\big)\big)(x,x) \notag
\end{align}
for both boundary conditions ${\rm X} = {\rm D}$ and ${\rm X} = {\rm N}$, and $ \mathbbm{P} $-almost all $ \omega \in \Omega $.
The next theorem shows that in the infinite-volume limit $ N_{\Lambda,{\rm X}}^{(\omega)} $ coincides with the
above-defined ensemble average (\ref{Def:N}), and therefore becomes 
independent of $ {\rm X} $ and $ \mathbbm{P} $-almost all $ \omega $.
\begin{theorem}\label{Thm:IDOS}
  Let $ V $ be a random potential in the sense of Definition~\ref{Def:pot}. 
  Moreover, let $\Lambda \subset \mathbbm{R}^d$ stand for bounded open cubes 
  centred at the origin. Then 
  there is a set $ \Omega_0 \in \mathcal{A} $ of full probability, 
  $ \mathbbm{P}\left( \Omega_0 \right) = 1 $, such that
  \begin{equation}\label{Eq:IDOSconv}
    N(E) =  \lim_{\Lambda \uparrow \mathbbm{R}^d} \, N_{\Lambda,{\rm X}}^{(\omega)}(E) 
  \end{equation}
  holds for  both  boundary conditions ${\rm X} = {\rm D}$
  and ${\rm X} = {\rm N}$, all $ \omega \in \Omega_0 $ and 
  all energies $ E \in \mathbbm{R} $ 
  except for the (at most countably many) discontinuity points of $ N $.
\end{theorem}
By a suitable ergodic theorem \cite{Kre85} 
the \emph{existence} and almost-sure \emph{non-randomness} of both infinite-volume limits in (\ref{Eq:IDOSconv}) 
are basically due to the assumed  $\mathbbm{R}^d $-ergodicity of $ V $.  
Under slightly different hypotheses the actual proof was
outlined in \cite{Mat93}. It uses functional-analytic arguments
first presented in \cite{KirMar82} for the case $ A = 0 $.
A different approach to the existence of these limits for $ A \neq 0 $,
using Feynman-Kac(-It{\^o}) functional-integral representations of
Schr{\"o}dinger semigroups \cite{Sim82,BrHuLe00}, 
can be found in \cite{Uek94,BrHuLe93}.
It dates back to \cite{Pas71,Nak77} for the case $ A = 0 $ and,
to our knowledge, works straightforwardly in the case $ A \neq 0 $ 
for $ \rm X = \rm D $ only. 
For $ A \neq 0 $ \emph{uniqueness} of the infinite-volume limit in
(\ref{Eq:IDOSconv}), that is, its independence of the boundary
condition $ \rm X $ (previously claimed without proof in \cite{Mat93}) follows 
from \cite{Nak00} if  
the random potential $ V $ is bounded and from \cite{DoIwMi01,HuSi02} if $ V $ is
bounded from below. 
For random potentials $ V $ yielding Schr\"odinger operators $ H(A,V) $ which are almost surely unbounded from below,
the proof of (\ref{Eq:IDOSconv}) can be found in \cite{HuLeMuWa01b,BrMuLe01}, 
see also \cite[Thm.~5.20]{PaFi92} for $ A = 0 $. 

\begin{remark}
      Similar as in equilibrium statistical-mechanics \cite{Rue99} there are more 
      general sequences of regions expanding to $ \mathbbm{R}^d $ than concentric open cubes $ \Lambda $ 
      for which (\ref{Eq:IDOSconv}) is true, see for example \cite[p.~105]{PaFi92}, \cite[p.~304]{CaLa90} or 
      \cite{DoIwMi01}. 
      Moreover, the convergence (\ref{Eq:IDOSconv}) holds for any other boundary condition $ \rm X $ for which the self-adjoint operator
      $ H_{\Lambda,{\rm X}}(A,V^{(\omega)}) $ obeys the inequalities 
      $ H_{\Lambda,{\rm N}}(A,V^{(\omega)}) \leq H_{\Lambda,{\rm X}}(A,V^{(\omega)}) \leq H_{\Lambda,{\rm D}}(A,V^{(\omega)}) $ 
      in the sense of (sesquilinear) forms. 
      The case of those mixed (in other words: Robin) boundary conditions, 
      which cannot be sandwiched between Dirichlet and Neumann boundary conditions, 
      is treated in \cite{Min03}.
\end{remark}

\section{Results in case of Gaussian random potentials}\label{Sec:Gauss}
This section compiles rigorous results on Lifshits tails, the density of states and spectral as well as dynamical localization in
case of Gaussian random potentials. 
The corresponding theorems are formulated under increasingly stronger conditions on the covariance function.

\subsection{Lifshits tails}
Since Gaussian random potentials $V$ have unbounded negative fluctuations, it is not surprising
that the leading low-energy fall-off of the integrated density of states is also Gaussian, even in the presence of a 
magnetic field. 
In particular, this type of fall-off ensures that $ H(A, V) $, although the latter is almost surely unbounded from below, 
may serve as the one-particle Schr\"odinger operator of a macroscopic system of non-interacting fermions 
in a random medium 
with well-defined specific free energy (\ref{eq:free}) and related thermodynamic quantities.
\begin{theorem}
\label{Thm:LifGaus}
Let $ V $ be a Gaussian random potential with covariance function $ C $. Then the leading
low-energy fall-off of the integrated density of states $ N $ is Gaussian in the sense that
\begin{flalign}\label{eq:LifGaus}
 &&
  \log N(E) \sim - \frac{E^2}{2 C(0)}  &&& (E \to - \infty).
\end{flalign}
$\mbox{}$
\end{theorem}
Theorem~\ref{Thm:LifGaus} dates back to Pastur \cite{Pas72,Pas77}, see also \cite{Nak77,KirMar82},  
and \cite{BrHeLe89a,BrHuLe93,Mat93,Uek94} for the magnetic case, where the last two references even allow the presence 
of certain \emph{random} magnetic fields.
The by now standard way of proving~(\ref{eq:LifGaus}), already used by Pastur, 
is in the spirit of (\ref{Eq:Theophys}) (setting there $ E_0 = - \infty $). One first determines 
the leading asymptotic behaviour of the Laplace-Stieltjes transform,
$ \widetilde N(\tau) := \int_{\mathbbm{R}} \d N(E) \, \exp(-\tau E) $, $ \tau > 0 $, of $ N $
as $ \tau \to \infty $ and
then applies an appropriate Tauberian theorem \cite{BGT89}.

The Lifshits tail (\ref{eq:LifGaus}) in case of Gaussian random potentials is highly universal. 
It only depends on the single-site variance $ C(0) = \mathbbm{E}\left[ V(0)^2 \right] > 0 $, but not 
on further details of the covariance function, the space dimension or the magnetic field. 
As physical heuristics and formula (\ref{Eq:Theophys}) already suggest, the 
(non-negative) kinetic-energy operator $ H(A,0) $ becomes irrelevant at extremely negative energies and the tail (\ref{eq:LifGaus}) 
is purely classical in the sense that $  \log N(E) \sim \log N_c(E) $ as $ E \to - \infty $.\\

In contrast to the universal classical Lifshits tail (\ref{eq:LifGaus}) 
of $ N $, the analogous tail
of the restricted integrated density of states $ R_l $ exhibits non-universal quantum 
behaviour in that it depends on the magnetic field and on details of the covariance function.
\begin{theorem}[\cite{BrHeLe91}]\label{Thm:LifRes}
  Let $ d = 2 $ and $ B > 0 $. Suppose that $ V $ is a Gaussian random potential with covariance function $ C $. 
  Moreover, let $ \sigma_l^2 > 0 $, see \eqref{Eq:2mom}. Then the  leading
  low-energy fall-off of the restricted integrated density of states 
  $ R_l $ is Gaussian in the sense that
  \begin{flalign}\label{Eq:LifRes}
    && \log R_l (E) \sim - \frac{E^2}{2 \Gamma_l^2} &&& (E \to - \infty).
  \end{flalign}
  Here the \emph{fall-off energy} $ \Gamma_l > 0$ is given by a solution of the
  maximization problem
  \begin{equation}\label{Eq:DecEn}
    \Gamma_l^2  := \!\! \sup_{\substack{\varphi \in P_l L^2({\mathbbm{R}}^2)
        \\ \langle \varphi , \varphi \rangle= 1}}  \int_{{\mathbbm{R}}^2} \! \! \! \! {\rm d} x  
    \int_{{\mathbbm{R}}^2} \! \! \! \! {\rm d} y   \left| \varphi(x) \right|^2  
    C(x-y) \, \left| \varphi(y) \right|^2.
  \end{equation}
  $\mbox{}$
\end{theorem}

    A proof of Theorem~\ref{Thm:LifRes} follows the lines of reasoning in \cite{BrHeLe91}, which
    amounts to establish the appropriate version of (\ref{Eq:Theophys}).
    The symmetry $ R_l(E) = 1 - R_l(-E) $, for all $ E \in \mathbbm{R} $, then immediately gives the high-energy growth 
    $ \log \big( 1 - R_l(E) \big) \sim - E^2/(2 \Gamma_l^2) $ as $ E \to \infty $. 
    For the Gaussian covariance (\ref{Eq.Gausscov}) a maximizer in (\ref{Eq:DecEn}) is given by 
    $ \varphi(x)   =  \sqrt{B /(l! \, 2 \pi )} \, \big[ \sqrt{B / 2 } \, (x_1 - \i x_2 )\big]^{l} \exp( - B \, | x |^2 / 4 ) $ 
    and the squared fall-off energy is explicitly found to be 
    \begin{equation}\label{eq:decayGauss}
      \Gamma^2_l = \left[ B\lambda^2 / ( B\lambda^2 + 1) \right]\, w_l(B \lambda^2 + 1),
    \end{equation}
    see \cite{BrHeLe91} and (\ref{Eq:levelwidth}), and also \cite{Ape87} for $ l = 0 $.  
    For a comparison of the fall-off energies in (\ref{eq:LifGaus}) and (\ref{Eq:LifRes}), 
    we offer the chain of inequalities 
    $ \sigma_l^4 /C(0) \leq \Gamma_l^2 \leq \sigma^2_l\leq C(0) $ which is actually valid \cite{BrHeLe91} for 
    the covariance function of a general $ \mathbbm{R}^2 $-homogeneous random potential,
    not only of a Gaussian one. 
%
\subsection{Existence of the density of states}
The continuity and non-negativity of the covariance function of a Gaussian random potential already imply
that the corresponding integrated density of states $ N $ is locally Lipschitz continuous, equivalently, that $ N $ 
is absolutely continuous on any bounded interval
and its (Lebesgue-) derivative $ D(E) = \d N(E)/\d E$, the \emph{density of states}, is locally bounded.
\begin{theorem}[\cite{FiHu97b,HuLeMuWa01a}]\label{Thm:DOS}
  Let $ V $ be a Gaussian random potential with non-negative covariance function $ C $.
  Then the integrated density of states $ N $ is locally Lipschitz continuous and the inequality 
  \begin{equation}
    \frac{\d N(E)}{\d E } \leq W(E) 
  \end{equation}
  holds for Lebesgue-almost all energies $ E \in \mathbbm{R} $ with some   
  non-negative $ W \in {\mathrm{L}}^\infty_{\mathrm{loc}}(\mathbbm{R}) $, which is independent
  of the magnetic field.
\end{theorem}

\begin{figure}[tbh]
\vspace*{0.2cm}
\begingroup\makeatletter%
\gdef\SetFigFont#1#2#3#4#5{%
  \reset@font
  \fontsize{8}{#2pt}
  \fontfamily{#3}\fontseries{#4}\fontshape{#5}%
  \selectfont}%
\endgroup%
\begin{center}
\begin{picture}(0,0)%
\includegraphics{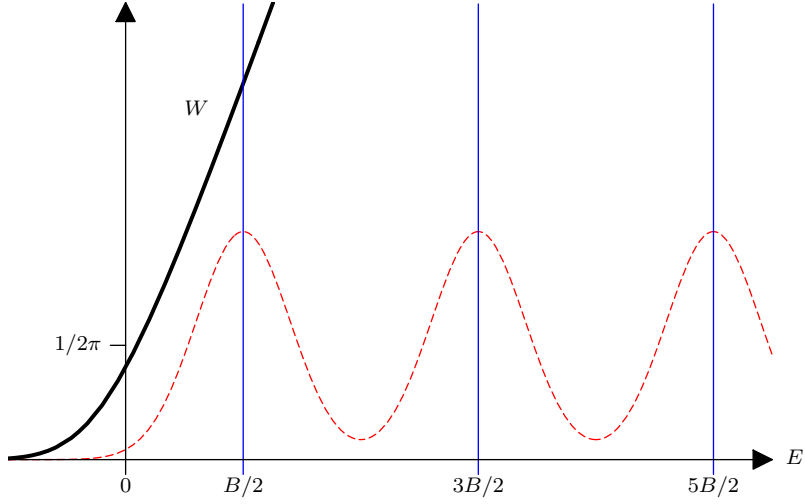}%
\end{picture}%
\setlength{\unitlength}{4144sp}%
\begingroup\makeatletter\ifx\SetFigFont\undefined%
\gdef\SetFigFont#1#2#3#4#5{%
  \reset@font\fontsize{#1}{#2pt}%
  \fontfamily{#3}\fontseries{#4}\fontshape{#5}%
  \selectfont}%
\fi\endgroup%
\begin{picture}(4680,2937)(429,-2536)
\put(1160,-2536){\makebox(0,0)[b]{\smash{\SetFigFont{8}{9.6}{\familydefault}{\mddefault}{\updefault}
$0$
}}}
\put(1864,-2536){\makebox(0,0)[b]{\smash{\SetFigFont{8}{9.6}{\familydefault}{\mddefault}{\updefault}
$B/2$
}}}
\put(3271,-2536){\makebox(0,0)[b]{\smash{\SetFigFont{8}{9.6}{\familydefault}{\mddefault}{\updefault}
$3B/2$
}}}
\put(4677,-2536){\makebox(0,0)[b]{\smash{\SetFigFont{8}{9.6}{\familydefault}{\mddefault}{\updefault}
$5B/2$
}}}
\put(5109,-2371){\makebox(0,0)[lb]{\smash{\SetFigFont{8}{9.6}{\familydefault}{\mddefault}{\updefault}
$E$
}}}
\put(1511,-281){\makebox(0,0)[lb]{\smash{\SetFigFont{12}{14.4}{\familydefault}{\mddefault}{\updefault}
$W$
}}}
\put(1016,-1686){\makebox(0,0)[rb]{\smash{\SetFigFont{8}{9.6}{\familydefault}{\mddefault}{\updefault}
$1/2\pi$
}}}
\end{picture}
\caption{Plot of an upper bound $ W(E) $ on $ D(E) $
  as a function of the energy $E$.
  Here $ D(E) = \d N(E)/\d E $ is 
  the density of states of the Landau Hamiltonian 
  with a Gaussian random potential with Gaussian covariance function (\ref{Eq.Gausscov}). 
  The dashed line is a plot of the graph of an approximation to $ D(E) $. 
  When zooming in, its bell-shaped parts near the Landau-level energies exhibit asymmetries. 
  The exact $ D(E)  $ is unknown. 
  Vertical lines indicate the delta peaks which reflect the 
  non-existence of the density of states 
  without random potential, $ V = 0 $.
  The single-step function $ \Theta(E) / 2\pi $ (not shown) 
  is the free density of states characterized by 
  $ V = 0 $ and $ B = 0 $, confer (\ref{DefclIDOS}). }\label{bild}
\end{center}
\end{figure}

A simple, but not optimal choice for the Lipschitz constant is given by
\begin{equation}
  W(E) = \left( r^{-1}   +  (2 \pi \tau )^{-1/2} \right)^{d} \; \frac{\exp\!\left\{ \tau E 
      + \tau^2 C(0) \left[ 1 - \varkappa_r^2/2 \right] \right\}}{\varkappa_r  \, \sqrt{2 \pi C(0)} }.
\end{equation}
Here $ r, \tau \in ] 0, \infty [ $ are arbitrary except that 
$ \varkappa_r := \inf_{| x | \leq r/\sqrt{d}} \, C(x) / C(0) > 0 $ must be strictly positive. 
By the assumed continuity of the covariance function, 
the latter condition is fulfilled at least for all sufficiently small $ r $. 
Figure~\ref{bild}, which is taken from \cite{HuLeMuWa01a}, contains the graph of $ W $ (after a certain numerical minimization) 
for the case of the Gaussian covariance function (\ref{Eq.Gausscov}) and $ d = 2 $.
The upper bound reveals that 
the density of states has no 
infinities for arbitrarily weak disorder, that is, for arbitrarily small $ C(0) > 0 $.\\

Theorem~\ref{Thm:DOS} relies on a so-called \emph{Wegner estimate} \cite{Weg81}, which bounds the average number of
eigenvalues of the finite-volume random Schr\"odinger operator $ H_{\Lambda,{\rm X}}(A,V) $ 
in a given energy interval $ [E_1 , E_2[ \, \subset \mathbbm{R} $ from above. 
More precisely, under the assumptions of Theorem~\ref{Thm:DOS} one has \cite{FiHu97b,HuLeMuWa01a} the inequality
\begin{equation}\label{Eq:Wegner}
  \mathbbm{E}\left[ N_{\Lambda,{\rm X}}(E_2) - N_{\Lambda,{\rm X}}(E_1) \right] 
  \leq  W(E) \, | E_2 - E_1 | 
\end{equation}
for both boundary conditions $ {\rm X} = {\rm D} $ and $ {\rm X } = {\rm N} $, and 
all $ E \geq E_2 $, provided the bounded open cube $ \Lambda \subset \mathbbm{R}^d $ is not too small.
The proof of (\ref{Eq:Wegner}) uses the one-parameter decomposition 
$ V^{(\omega)}(x) =: V^{(\omega)}_0(x) + V^{(\omega)}(0) \, C(x)/ C(0)$ of 
the Gaussian random potential $ V $ with covariance function $ C $.
By this definition $ V_0 $ is a non-homogeneous Gaussian random
field which is stochastically independent of the Gaussian random variable $ V(0) $.
A second essential ingredient of the proof is the abstract one-parameter spectral-averaging estimate 
of Combes and Hislop \cite[Cor.~4.2]{CoHi94} (for an equivalent formulation see \cite[Lemma~1]{HuLeWa01}). 
Finally, the independence of $ W $ of the magnetic field arises from the diamagnetic inequality for
Neumann partition functions~\cite[App.~A]{HuLeMuWa01a}, see also \cite{HuSi02}.\\

The analogue of Theorem~\ref{Thm:DOS} for the single-band approximation of the random Landau Hamiltonian yields 
the existence and boundedness of the probability density
$\d R_{l}(E)/\d E $, the \emph{restricted density of states}.  
\begin{theorem}[\cite{HuLeWa01}]
  Let $ d = 2 $ and $ B > 0 $. Suppose that $ V $
  is a Gaussian random potential with non-negative covariance function $ C $.
  Then the restricted integrated density of states $ R_{l} $ is Lip\-schitz continuous and the inequality 
  \begin{equation}\label{Eq:restdos}
    \frac{\d R_{l}(E)}{\d E } \leq \frac{1}{\sqrt{2 \pi \Gamma_l^2}}
  \end{equation}
  holds for Lebesgue-almost all energies $ E \in \mathbbm{R} $, where the constant 
  $ \Gamma_l $ is the fall-off energy of the $ l $th Landau band,
  see (\ref{Eq:DecEn}).
  If additionally $ C $ is circularly symmetric,
  one also has  
  \begin{equation} 
    \frac{\d R_{l}(E)}{ \d E } \leq \frac{C(0)}{\sigma^2_l}  \frac{1}{\sqrt{2 \pi C(0)}} \, \exp\left( - \frac{E^2}{2 C(0)} \right),
  \end{equation}
  where $ \sigma_l $ is the width of the $l$th Landau band, see \eqref{Eq:2mom}.
\end{theorem}
    In the physics literature \cite{ZiLa66,Efe97,Sus98} one often considers the limit of a  \emph{delta-correlated}
    Gaussian random potential informally characterized by  $C(x) = u^2 \, \delta(x) $ 
    with some constant $ u > 0 $. 
    It emerges from the Gaussian random potential with the Gaussian covariance function 
    (\ref{Eq.Gausscov}) by choosing $ C(0) = u^2 (2 \pi \lambda^2)^{-d/2} $ and letting $ \lambda \downarrow 0 $. 
    In particular,  several simplifications occur
    in this limit for the random Landau Hamiltonian ($ d = 2 $) in its single-band approximation. 
    For example,  the band width becomes independent of 
    the Landau-level index, $ \sigma_l^2 = \sigma_0^2 = u^2 B/(2\pi) $, and the squared 
    fall-off energy (\ref{eq:decayGauss}) takes the form
    $ \Gamma_l^2 = \left(u^2 B /4 \pi \right) (2 l)!/(l ! \, 2^l)^2 $.
    More interestingly, there is an explicit expression for $ \d R_0(E)/\d E $ in this limit and also for
    $ \d R_l(E)/\d E $, if the subsequent  
    high Landau-level limit $ l \to \infty $ is performed.
    The first result is due to Wegner \cite{Weg83} and reads 
    \begin{equation}\label{Eq:Wegnerdelta}
      \frac{\d R_0(E)}{ \d E } = 
      \frac{2}{\pi^{3/2} \sigma_0}  \,
      \frac{\exp(\eta^2)}{1 + \left[2 \pi^{-1/2} 
          \int_0^\eta \! {\rm d}\xi \, \exp(\xi^2) \right]^2}, 
       \qquad \eta := \frac{E}{ \sigma_0},
     \end{equation} 
     see also \cite{BeGrIt84,KlPe85,MaPu92}.
     As for the second result, it is known \cite{BenCha86,Sal87} that $\d R_l(E)/\d E $ 
     becomes semi-elliptic, 
     \begin{equation}\label{eq:hll}
       \lim_{l \to \infty} \frac{\d R_{l}(E)}{ \d E } = \frac{1}{2 \pi \sigma_0} \, 
       \Theta\left( 4 - \eta^2 \right) \, \sqrt{4  - \eta^2}.
     \end{equation}
     While the right-hand side of (\ref{Eq:restdos}) remains finite in the delta-correlated limit, it   
     diverges asymptotically as $ l^{1/4}/(\pi^{1/4}\sigma_0) $ in the additional limit $l \to \infty$, 
     which is in accordance with~\cite{Ben87}.

   The existence of a bounded restricted density of states $\d R_0(E)/\d E $ in the 
   delta-correlated limit of a Gaussian random potential  
   stands in contrast to certain situations with random \emph{point} impurities. 
   In all of the considered cases \cite{BeGrIt84,DoMa97,PuSc97,Erd98,Scr99}, it has been shown
   that the restricted random operator 
   $ P_0 V P_0 $ has almost surely an infinitely degenerate eigenvalue, if the magnetic-field strength 
   $B$ is sufficiently large; see also (\ref{eq:BGI}) below.
   Part of these results have been unified in \cite{CLP01}. %

\subsection{Spectral and dynamical localization}
Since even weak disorder should be able to suppress quantum-mechanical tunnelling at sufficiently negative energies, 
bound states should emerge in such energy (or disorder) regimes in case of 
Gaussian random potentials. 
Therefore, the content of the subsequent theorems is often taken for granted in the physics literature. 
To our knowledge however, Theorem~\ref{Thm:loc} and Theorem~\ref{thm:ueki} contain the first rigorous results on localization 
by an $ \mathbbm{R}^d $-ergodic random potential in the multi-dimensional case $ d \geq 2 $. 
For their precise formulations 
we first need to define a property the covariance function~$ C $ may have.

\begin{definition}\label{Def:R}
  A covariance function $ C: \mathbbm{R}^d \to \mathbbm{R} $ 
  of a (Gaussian) random potential is said to have \emph{property~\ass{R}} if 
  it admits the representation 
  $  C(x) = \int_{\mathbbm{R}^d}\! \d y \, \gamma(x + y) \, \gamma(y) $
  for all $ x \in \mathbbm{R}^d $ with some non-zero $ \gamma \in {\rm L}^2(\mathbbm{R}^d) $  which is both
\begin{itemize} 
  \item[(i)]
    H\"older continuous of order $ \alpha $, that is,  
    there exist constants $ \alpha \in ]\, 0, 1] $ and $ a > 0 $
    such that 
    $ | \gamma(x+y)  - \gamma(x) | \leq a | y |^\alpha $  
    for all $ x \in \mathbbm{R}^d $ and all $ y $ in some neighbourhood of the origin $ 0 \in \mathbbm{R}^d $, and
  \item[(ii)] 
    non-negative and has sufficiently rapid decay at infinity in the sense that
    $ 0 \leq \gamma(x) \leq \gamma_0 \, ( 1 + | x |)^{-\beta} $ for Lebesgue-almost all $ x \in \mathbbm{R}^d $ 
    with some constant 
    $ \gamma_0 > 0 $ and some exponent $ \beta > 13 \, d / 2 + 1 $.
\end{itemize}
\end{definition}
\begin{remarks}
  \begin{nummer}
    \item\label{Remark4.1(ii)}
      In accordance with the Wiener-Khintchine criterion, 
      the representation in Definition~\ref{Def:R} (without requiring~(i) and~(ii)) 
      is equivalent to the assumption that the (non-negative) 
      spectral measure of  $ V $ is absolutely continuous with respect to Lebesgue measure.  
      A prominent example having property \ass{R}  
      is the Gaussian covariance
      function (\ref{Eq.Gausscov}) with arbitrary correlation length $ \lambda > 0 $ for which 
      $ \gamma(x) = \big(2/ (\pi \lambda^2)\big)^{d/4} \sqrt{C(0)} \, \exp\left( - |x|^2/ \lambda^2\right) $.
    \item
      Property \ass{R} in particular ensures the non-negativity of $ C $. 
      It also implies \cite{Fer75} 
      the  $\mathbbm{P}$-almost sure continuity, and hence local
      boundedness, of  the realizations of $V$. The decay of $\gamma$ at
      infinity yields the mixing property of $V$, hence ergodicity.
      In fact, property \ass{R} requires a compromise between local
      dependence and global independence of $V$.
      From a physical point of view, both requirements are plausible:
      The effective one-particle potential $ V $
      should be smooth due to screening. By the same token it is expected
      that impurities only weakly influence each other over long distances.
   \end{nummer}
\end{remarks}

\begin{theorem}[\cite{FiLeMu00}]\label{Thm:loc} 
  Let $ V $ be a Gaussian random potential with a covariance function 
  having property~\ass{R}. Then
  \begin{itemize}
    \item[(i)]
      for every coupling constant $ \zeta > 0 $ there exists an energy $ E_{pp} < 0 $ such
      that the spectrum of $ H(A, \zeta V^{(\omega)}) $ is only pure point in the half-line $ ] - \infty , E_{pp} ] \, $
      for $ \mathbbm{P} $-almost all $ \omega  \in \Omega $.
    \item[(ii)]
      for every energy $ E < 0 $ there exists a coupling constant $ \zeta_{pp} > 0 $ such that
      for every $ \zeta \in \, ] \, 0 , \, \zeta_{pp}] $ there exists $ \Omega_\zeta \in \mathcal{A} $ 
      with $ \mathbbm{P}(\Omega_\zeta) = 1 $ and the spectrum of $ H(A,\zeta V^{(\omega)}) $ 
      is only pure point in $ ] - \infty , E] \, $ 
      for all $ \omega \in \Omega_\zeta $.
  \end{itemize}    
\end{theorem} 
Given the Wegner estimate (\ref{Eq:Wegner}), the proof of Theorem~\ref{Thm:loc} 
is based on a so-called \emph{multi-scale analysis} in the spirit of the fundamental work of 
Fr\"ohlich and Spencer \cite{FrSp83}. They gave the first rigorous result on localization in case of a multi-dimensional lattice model,
the original one due to Anderson \cite{And58}.
The multi-scale analysis invokes elements from Kolmogorov-Arnold-Moser theory 
for coping with small denominators in order to bound resolvents of finite-volume random Schr\"odinger operators
with high probabilities. Its applicability to the present situation requires additional assumptions on the covariance function 
beyond those needed for the Wegner estimate. 
The technical realization of the proof of Theorem~\ref{Thm:loc} is patterned after
the ``fixed-energy'' analysis of von Dreifus and Klein \cite{DrKl91}
in order to handle the long-ranged correlations of the fluctuations of Gaussian random potentials.
In addition to that, different random potentials are used on
different length scales. The idea behind this is to replace the given
long-ranged correlated random potential $V$ on the length scale $\ell_{n}$
by the element $V_{n}$ of a sequence $\{V_{n}\}_{n\in\mathbbm{N}}$ of
(Gaussian) random potentials such that ~(i)~ $\{V_{n}\}$ converges to $V$ in a
suitable sense as $ n \to \infty $ and that ~(ii)~ each $V_{n}$ has 
finite-ranged correlations, but
with a spatial extent that grows with increasing $n$. 
For the adaptation to the continuous space $ \mathbbm{R}^d $ the proof follows 
the lines of Combes and Hislop \cite{CoHi94} and Figotin and Klein \cite{FiKl96}. 

\begin{remarks}
  \begin{nummer}
    \item
      The assumptions of Theorem~\ref{Thm:loc} also allow for so-called deterministic
      random potentials, as is illustrated by the Gaussian covariance (\ref{Eq.Gausscov}) for one space dimension.
    \item
      It would be interesting to see whether the proof of Theorem~\ref{Thm:loc} 
      can be simplified by adapting the continuum-extension \cite{Aiz02}
      of the fractional-moment approach by Aizenman and Molchanov \cite{AiMo93,Aiz94,AiGr98,AiSc00,AiSc01}
      to (spectral) localization, which was 
      originally developed for the lattice~$ \mathbbm{Z}^d $.
  \end{nummer}
\end{remarks}

As a consequence of the unbounded negative fluctuations of Gaussian random potentials the authors of \cite{FiLeMu00}
were only able to prove algebraic instead of exponential decay of the eigenfunctions corresponding to 
the pure-point spectrum. This technical problem should be mastered by extending either the 
"variable-energy" multi-scale analysis \cite{Spe88,DrKl89,KiSt98} 
or the powerful bootstrap programme of Germinet and Klein \cite{GeKl01,GeKl02} 
to certain Schr\"odinger operators which are almost surely unbounded from below. Indeed, by doing the latter, 
Ueki \cite{Uek02} succeeded in showing
exponential localization for certain Gaussian random potentials provided their covariance function is compactly supported. 
Along the same lines, he obtained first results on dynamical localization.
The following theorem may be deduced by specializing his results.
\begin{theorem}[cf.~\cite{Uek02}]\label{thm:ueki}
  Let $ V $ be a Gaussian random potential with covariance function $ C $ having property~\ass{R}. Additionally assume 
  that $ C $ is compactly supported. Then there exists an energy $ E_{pp} < 0 $ such that
  \begin{itemize}
    \item[(i)] for
      $ \mathbbm{P} $-almost all $ \omega  \in \Omega $ 
      the spectrum of $ H(A, V^{(\omega)}) $ in the half-line $]-\infty , E_{pp} ] $ is only pure point
      with exponentially localized eigenfunctions. 
  \item[(ii)]
    for every $ E \in ]-\infty , E_{pp} [ $ there exists
    $ \epsilon > 0 $ such that strong dynamical localization holds in the energy interval $ I:=]E-\epsilon,E+\epsilon[ $ in the sense 
    that 
    \begin{equation}
      \mathbbm{E}\left[ \, \sup_{t \in \mathbbm{R}} \, 
        \int_{\mathbbm{R}^d} \!\! \d x \; \big| \psi_{t,I}(x) \big|^2 \, | x |^q \right]
      < \infty
    \end{equation}\label{eq:dynloc}
    for all $ q \geq 0 $ and all $ \psi_0 \in \mathcal{C}^\infty_0(\mathbbm{R}^d) $. 
    {\rm [}Recall the definition of $ \psi_{t,I}^{(\omega)} $
    below \eqref{eq:dyn}.{\rm ]}
 \end{itemize}
\end{theorem}

\begin{remark}
      As is shown in \cite{Uek02} the assumptions of Theorem~\ref{thm:ueki} even imply dynamical localization in the (strong) Hilbert-Schmidt sense.
      In case of $\mathbbm{Z}^d$-ergodic random potentials dynamical localization 
      has been proven in various forms \cite{FrSp83,MaHo84,GeDBi98,AiGr98,BaFi99,DaSt01,Sto01,GeKl01,GeKl02b}
      under practically no further assumptions than required to prove spectral localization.
      It would be desirable to prove some sort of dynamical localization also in the situation of Theorem~\ref{Thm:loc}. 
\end{remark}  

\section{Results in case of Poissonian random potentials}\label{Sec:Poisson}
In comparison to Gaussian random potentials, less is known about regularity properties of the 
integrated density of states $ N $ and localization in case of Poissonian random potentials for arbitrary space dimension.
Most results concern the Lifshits tail of $ N $. 
\subsection{Lifshits tails}
In contrast to the case of Gaussian random potentials, there is a huge multiformity of Lifshits tails in the Poissonian case.

For vanishing magnetic field and non-negative Poissonian random potentials with rapidly decaying single-impurity potentials,
the leading low-energy fall-off of $ N $ was first identified by Lifshits \cite{Lif63,Lif64}. 
Using the strategy already described after Theorem~\ref{Thm:LifGaus}, a rigorous proof \cite{Pas77,Nak77} 
relies on Donsker and Varadhan's celebrated 
large-deviation result~\cite{DonVar75} on the long-time asymptotics of certain Wiener integrals. 
We summarize the rigorous version of Lifshits' result together with the corresponding one of Pastur \cite{Pas77} for slowly decaying
single-impurity potentials in 
\begin{theorem}[\cite{DonVar75,Pas77,Nak77}]\label{Thm:Lif}
  Let $ V $ be a non-negative Poissonian random potential with mean concentration $ \varrho > 0 $.  
  Furthermore, assume that the single-impurity potential $ U \geq 0$ has one of the following two decays at infinity: 
  \begin{indentnummer*}
  \item[\ass{D1}] $ U $ is 
    compactly supported or has a decay more rapid than algebraic with exponent $ d + 2$ in the sense that   
    $ \limsup_{| x | \to \infty} |x|^{d+2}\,  U(x) = 0 $,
  \item[\ass{D2}] $ U $ has (definite) algebraic decay  with exponent $ \alpha \in ] \, d,  d + 2 \, [$ in the sense that   
    $ \lim_{| x | \to \infty} |x|^{\alpha}\,  U(x) = g $ with some constant $ g > 0 $.
  \end{indentnummer*}
  Moreover, assume that the magnetic field vanishes, $ (B_{ij}) = 0 $. 
  Then the leading low-energy fall-off of the integrated density of states reads
  \begin{flalign}\label{Eq:TailnoB}
    &&
    \quad \log N(E) \sim \left\{
      \begin{array}{c@{\qquad\mbox{for the decay}\;\;}l} 
        \displaystyle - \varrho \left(\frac{\kappa_d}{2 E}\right)^{d/2} & \mbox{\ass{D1}} \\[2ex] 
        \displaystyle - C_d(\alpha, \varrho) \left(\frac{g}{E}\right)^{d/(\alpha-d)}  
         & \mbox{\ass{D2}}
      \end{array}      
    \right\}
    &&& (E \downarrow 0).  
  \end{flalign}
  Here $ \kappa_d $ is the lowest eigenvalue of the negative Laplacian $ 2 H(0,0) $, when Dirichlet
  restricted to a ball in $ \mathbbm{R}^d $ of unit volume. Moreover, we have introduced the
  constant 
  $ C_d(\alpha, \varrho) := \frac{\alpha -d}{d}
  \left[ \varrho\,\frac{d}{\alpha} \, \frac{\pi^{d/2}}{\Gamma(1 + d/2)} \,  
    \Gamma\!\left(\frac{\alpha -d}{\alpha}\right)\right]^{\alpha/(\alpha-d)} $.
\end{theorem} 
\begin{remarks} 
  \begin{nummer} 
    \item 
      One has, for example, $ \kappa_1 = \pi^2 $, $ \kappa_2 = \pi a_0^2 $, 
      where $ a_0 = 2.4048\dots $ is the smallest positive zero of the zeroth Bessel function 
      of the first kind~\cite{GrRy}, and $ \kappa_3 = \pi^2 (4 \pi /3 )^{2/3} $. 
    \item 
      Convincing arguments for the
      validity of Lifshits' result (\ref{Eq:TailnoB}) for the decay \ass{D1} were also given in \cite{FrLu75,Lut76}.
      An alternative (rigorous) proof of the underlying long-time asymptotics 
      is due to Sznitman who invented a coarse-graining scheme called the 
      \emph{method of enlargement of obstacles} \cite{Szn98}. More elementary proofs which rely on 
      Dirichlet-Neumann bracketing were found in \cite{KirMar83,Sto99}, but for the price of obtaining 
      only the so-called
      \emph{Lifshits exponent} (here: $ d/2 $) and not the other constants in (\ref{Eq:TailnoB}) for the decay~\ass{D1}.
    \item
      As an aside, we note that (\ref{Eq:TailnoB}) for the decay \ass{D1} with $ d = 1 $ remains valid in the limiting 
      case of Poissonian point
      impurities, $ U(x) = u_0 \, \delta(x) $, $u_0 > 0 $; see \cite{Egg72,GrPa75,Kot76} and \cite[Thm.~6.7]{PaFi92}.
   \end{nummer}
\end{remarks}
For (non-negative) single-impurity potentials $ U $ with rapid decay \ass{D1}, the Lifshits tail  
is insensitive to the details of the decay of $ U $ and is dominated by the quantum kinetic energy.
It has therefore a quantum character.
Over against this, if $ U \geq 0 $ has the slow decay \ass{D2}, the Lifshits tail sensitively 
depends on the details of this decay.
It is classical in character in that 
$  \log N(E) \sim \log N_c(E) $ as $E \downarrow 0$.
Therefore, the character of the tail (\ref{Eq:TailnoB}) changes from purely quantum to purely classical,
when the decay changes from~\ass{D1} to~\ass{D2}.
The Lifshits tail for the 
borderline case of algebraic decay with exponent $ \alpha = d + 2 $ seems to be open. 
In view of (\ref{Eq:Theophys}) 
we have the following 
\begin{conjecture} 
If $ \lim_{| x | \to \infty} |x|^{d+2}\,  U(x) = g $ with some constant $ g > 0 $, then
\begin{flalign}
  &&
  \quad \log N(E) \sim - \left[ \varrho^{\frac{2}{d+2}} \left(\frac{\kappa_d}{2 E}\right)^{\frac{d}{d+2}}   
    + \big[C_d(d+2,\varrho)\big]^{\frac{2}{d+2}} \left(\frac{g}{E}\right)^{\frac{d}{d+2}}  \right]^{\frac{d+2}{2}} \!\!
  &&& (E \downarrow 0).
\end{flalign}
\end{conjecture}
This tail exhibits a mixed quantum/classical character.\\

A similar transition from a purely quantum to a purely classical Lifshits tail can be observed in case of 
the random Landau Hamiltonian with non-negative Poissonian potential. 
However, since the Landau Hamiltonian possesses ground-state (wave) functions with Gaussian decay,
the borderline decay of $ U $ turns out to be Gaussian and not algebraic (with exponent $ d + 2 = 4 $).

\begin{theorem}[\cite{BrHuLe95,Erd98,HuLeWa99,HuLeWa00,Erd00}]\label{Thm:LifB}
  Let $ d = 2 $ and $ B > 0 $. Suppose that $ V $ is a non-negative Poissonian random potential 
  with mean concentration $ \varrho > 0 $. Furthermore, assume that the 
  single-impurity potential $ U \geq 0 $
  has one of the following three decays at infinity: 
\begin{indentnummer*}
  \item[\ass{D3}] $ U $ is
   compactly supported or has super-Gaussian decay in the sense that 
    $ \limsup_{| x | \to \infty} | x |^{-2} \log U(x) = - \infty $. 
   \item[\ass{D4}] $ U $ has (definite) Gaussian decay in the sense that
    $ \limsup_{| x | \to \infty} | x |^{-2} \log U(x) $\hspace{0pt}$= - \lambda^{-2} $
    with some length $ \lambda > 0 $. 
  \item[\ass{D5}] $ U $ has sub-Gaussian decay in the sense that $ \limsup_{| x | \to \infty} | x |^{-2} \log U(x) = 0 $. 
    Moreover, the decay of $ U $ is
    integrable and regular in the sense of~~\cite[Def.~3.5]{HuLeWa00}.
\end{indentnummer*}
Then the leading low-energy fall-off of the integrated density of states reads
    \begin{flalign}\label{Eq:LifB}
      &&
      \quad \log N\!\left(\frac{B}{2}+E\right) \sim \left\{
      \begin{array}{c@{\quad\mbox{for the decay}\;\;}l}
        \displaystyle \frac{2 }{B} \, \pi \varrho \, \log E & \ass{D3} \\[2ex]
        \displaystyle
         \Big(\frac{2}{B}  + \lambda^2 \Big) \, \pi \varrho \, \log E & \ass{D4} \\[2.5ex]
        \displaystyle \log N_c(E) & \ass{D5} 
      \end{array}     
      \right\} \!\!\!
      &&& (E \downarrow 0). 
    \end{flalign}
    $\mbox{}$
\end{theorem}  
\begin{remarks} 
  \begin{nummer} 
    \item 
      For super-Gaussian decay~\ass{D3} and Gaussian decay~\ass{D4} the integrated density of states has a power-law fall-off
      (on a logarithmic scale). 
      The corresponding exponent $ 2 \pi \varrho / B $ in (\ref{Eq:LifB}) for the decay~\ass{D3} is just the mean number of 
      impurities in a disc of radius $ \sqrt{ 2 / B } $. 
      Two important examples for \ass{D5} are an algebraic decay with exponent $ \alpha \in ] \, 2 , \infty [ $
      (see Theorem~\ref{Thm:Lif}) and a stretched-Gaussian decay in the sense that 
      $ \lim_{|x|\to \infty} |x|^{-\beta} \log U(x) = - \lambda^{-\beta} $ with some exponent $ \beta \in ]\, 0, 2[ $ 
      and some decay length $ \lambda > 0 $.
      For stretched-Gaussian decay the explicit form 
      of (\ref{Eq:LifB}) 
      reads \cite{HuLeWa99}
      \begin{flalign}
        && \log N(B/2 + E ) \sim - \lambda^{2} \pi \varrho \,  | \log E |^{2/\beta} &&&
        (E \downarrow 0).
      \end{flalign}  
      For an algebraic decay the explicit form 
      of (\ref{Eq:LifB}) 
      coincides with the right-hand side of 
      (\ref{Eq:TailnoB}) for the decay~\ass{D2} with $ d = 2 $, even if $ \alpha \geq 4 \, ( = d + 2 ) $ \cite{BrHuLe95}.
      Other examples for~\ass{D5} causing more exotic fall-offs can be found in \cite{HuLeWa00}. 
   \item
     The hard part of the proof of Theorem~\ref{Thm:LifB} deals with the compactly supported case of~\ass{D3}. 
     It is due to Erd\H{o}s who developed a version \cite{Erd98} of 
     the method of enlargement of obstacles \cite{Szn98} which takes into account the presence of a magnetic field. 
     With this method he also confirmed  \cite{Erd00} the mixed quantum/classical character 
     of the Lifshits tail in case of the Gaussian decay~\ass{D4}, which was conjectured in~\cite{HuLeWa99}. 
   \end{nummer} 
\end{remarks} 
We have seen that the Gaussian Lifshits tail of the integrated density of states $ N $ of the random Landau Hamiltonian with 
a Gaussian random potential is slower than that of $ R_0 $, 
the integrated density of states of its lowest band (since $ \Gamma_0^2 < C(0) $ unless $ C(x) = C(0) $ for
all $ x \in \mathbbm{R}^2 $, which however is ruled out by ergodicity). 
Not unexpectedly, this ceases to be so in case of a Poissonian random potential with a non-negative
single-impurity potential. 
\begin{theorem}[\cite{War01}]\label{ThmRest0}
  Assume the situation of Theorem~\ref{Thm:LifB}. Furthermore, let the single-impurity potential be locally bounded, 
  $ U \in {\rm L}^\infty_{\rm loc}(\mathbbm{R}^2) $. 
  Then the leading low-energy fall-off of the restricted integrated density of states $ R_0 $  
  to zero near 
  zero energy coincides with that of $ N $ near $ B/2$ , that is
  \begin{flalign}\label{Eq:Restr0}
    && \log R_0(E ) \, \sim \, \log N\!\left(\frac{B}{2} + E \right)  
    &&& (E \downarrow 0).
  \end{flalign} 
  $\mbox{}$
\end{theorem}

    The proof of Theorem~\ref{ThmRest0} follows from Theorem~\ref{Thm:LifB} and 
    a two-sided estimate on the Laplace-Stieltjes transform $ \widetilde R_0  $ of $ R_0 $.
    For the lower estimate see \cite[Eq.~(3.7)]{BrHuLe95}. The upper estimate,  
    $ \widetilde R_0 (\tau) \leq   \e^{\tau B/2} \, \widetilde N(\tau) \; 2 \pi/ B$, stems from the Jensen-Peierls inequality
    applied to the right-hand side of \cite[Eq.~(3.7)]{HuLeWa99}.
    For the case of an algebraic decay, (\ref{Eq:Restr0}) has been given already in \cite{BrHuLe95},
    and for compactly supported $ U $ implicitly in \cite{Erd98}.

\begin{remark}
    Some results on Lifshits tails of higher Landau bands ($ l \geq 1$)
    are available in \cite{BrHuLe95,HuLeWa99}.
    The leading high-energy growth of $ R_l $ (for any $ l \geq 0 $) corresponding to a non-negative single-impurity potential $ U $ 
    coincides with the leading low-energy fall-off of the $ l $th restricted integrated density of states corresponding to $ - U $,
    because $ P_l - P_l \Theta(E-P_l V P_l)P_l = P_l \Theta(P_l V P_l - E)P_l $.
\end{remark}  

For all three types of decay \ass{D3}--\ass{D5} considered in Theorems~\ref{Thm:LifB} and \ref{ThmRest0},
$ N $ is continuous at $ B/2 $ and hence $ R_0 $ is continuous at zero energy.
This stands in contrast to the case of Poissonian \emph{point}-impurities, $U(x) = u_0 \, \delta(x)$, $u_0 > 0$, 
for which Br\'ezin, Gross and
Itzykson \cite{BeGrIt84} managed to calculate $ R_0 $ exactly and explicitly  
by using (non-rigorous)
supersymmetric functional integration, see also \cite{KlPe85}.
For the leading low-energy fall-off they obtained
\begin{align}
  &\qquad \lim_{E \downarrow 0} \,\, R_0(E) = 
  1 - \frac{2 \pi \varrho}{B} && \text{if} 
  && \frac{2 \pi \varrho}{B} < 1, \label{eq:BGI} \\
  & \qquad \lim_{E \downarrow 0} \,\, \left| \log E \right| \, R_0(E) = 
  1 && \text{if} 
  && \frac{2 \pi \varrho}{B} = 1, \\
  & \qquad \lim_{E \downarrow 0} \, 
  \frac{\log R_0(E)}{\left| \log E \right|} 
  = 1 - \frac{2 \pi \varrho}{B}  && \text{if} 
  && \frac{2 \pi \varrho}{B} > 1. 
  \label{eq:BGI1} 
\end{align}
Erd\H{o}s \cite{Erd98} has given a rigorous proof that the size of the jump of $ R_0 $ at zero
energy in case $ 2 \pi \varrho / B  < 1 $ is not smaller than the right-hand side of (\ref{eq:BGI}).\\

We close this subsection with two complementary remarks.
\begin{remarks}
  \begin{nummer}
  \item
    First rigorous results on Lifshits tails in case of three space dimensions $ d = 3 $  
    and a constant magnetic field of strength $ B > 0 $ are available. 
    For super-Gaussian decay~\ass{D3}, Gaussian decay~\ass{D4} and stretched-Gaussian decay 
    the so-called Lifshits exponent is
    identified \cite{War01} to coincide in all three cases with the corresponding one for $ d = 1 $,
    \begin{equation}\label{eq:lifexp}
      - \lim_{E \downarrow 0} \frac{\log \, \left| \log N\big( B/2 + E \big) \right|}{\log E} = \frac{1}{2}.
    \end{equation}
     This may be ascribed to the 
     effective zero-field motion of the particle parallel to the magnetic field, which dominates the low-energy asymptotics. 
    Actually, in \cite{War01}  somewhat more detailed information on the fall-offs depending on the actual decay can be found.
    For example, for algebraic decay~\ass{D2} with $ d = 3 $, that is, with exponent $ \alpha \in ] \, 3, 5 [ $, 
    the tail coincides with the corresponding one for $ B = 0 $ (see Theorem~\ref{Thm:Lif}) and has therefore a classical
    character \cite{HuKiWa02}.
    For heuristic explanations of (\ref{Eq:LifB}) and their (conjectured) analogues for $ d = 3 $, see \cite{LeWa03}.
    \item
      For vanishing magnetic field, Lifshits tails have been 
      investigated also for non-positive single-impurity potentials, $ U \leq 0 $.
      Depending on whether $ U $ is singular or not, the corresponding
      tail exhibits a quantum or classical character. For details, see \cite{Pas77,LiGr88,PaFi92,PaKl98}.
      The results are again consistent with formula~(\ref{Eq:Theophys}) (setting there 
      $ E_0 = - \infty $). 
    \end{nummer}
\end{remarks}

\subsection{Existence of the density of states and spectral localization} 
There are only a few rigorous results on these issues for Poissonian random potentials. 
For a special class of non-negative single-impurity potentials $ U $, Tip \cite{Tip94} has proven that the integrated
density of states $ N $ is absolutely continuous at sufficiently high energies.
The only localization result is due to Stolz \cite{Sto95}. It concerns the case of one space dimension.  
\begin{theorem}[\cite{Sto95}]\label{Thm:stolz}
  Let $ d = 1 $ and let $ V $ be a non-negative Poissonian random potential.  
  Moreover, let the single-impurity potential $ U \geq 0 $ be 
  compactly supported.  
  Then the almost-sure spectrum $ [0, \infty[$ of $ H(0,V) $ is only pure point with exponentially localized
  eigenfunctions.
\end{theorem} 
The proof builds on techniques which are available only for one dimension and 
are nicely summarized and discussed in the recent survey \cite{Sto02}. 

\begin{remark}
To our knowledge, the only other rigorous works \cite{MaSc87,CoHi94} which, among the rest, 
deal with localization proofs for Poissonian random potentials 
in multi-dimensional situations have to assume an additional randomness 
of the impurity coupling-strengths.
\end{remark}
\section{Some open problems} 
While most rigorous works on random Schr\"odinger operators concern
$\mathbbm{Z}^d$-ergodic random potentials, the present survey  has
focused on $\mathbbm{R}^d$-ergodic ones. More precisely, for Gaussian
and Poissonian random potentials, rigorous results have been presented
on the integrated density of states and Anderson localization.
In this context, a lot of issues, which are interesting from the (theoretical-) physics point of view, are still unsolved.

One major open problem concerns a proof of Anderson localization 
in case of (non-negative) Poissonian random potentials for arbitrary space dimension.
Another problem is to isolate the weakest possible conditions for an $ \mathbbm{R}^d $-ergodic random potential 
which imply continuity of the corresponding integrated density of states if $ d \geq 2 $ and $ A = 0 $.
In particular, one may ask: Is $ \mathbbm{R}^d $-ergodicity already enough?

Most striking is definitely the fact that there is not a single non-zero  $ \mathbbm{R}^d $- or 
$ \mathbbm{Z}^d $-ergodic random potential
for which the existence of an absolutely continuous component in the energy spectrum has been proven, that is, spectral delocalization 
in certain energy regimes. After all, physical intuition, approximate calculations 
and numerical studies suggest the occurrence of a mobility edge if $ d \geq 3 $. 
If $ d = 2 $ it is not yet rigorously settled whether
the whole energy spectrum 
is almost surely only pure point or not.  

Last but not least, one can hardly claim to utterly understand electronic properties of disordered solids 
without having a more solid foundation of their transport theory.

\section*{Acknowledgement}
Our thanks go to Alexandra Weichlein for helpful remarks.
This work was partially supported by the Deutsche Forschungsgemeinschaft (DFG) under grant no.\ Le 330/12. 
The latter is a project within the DFG Priority Programme SPP 1033 
``Interagierende stochasti\-sche Systeme von hoher Komplexit\"at''.
Peter M\"uller acknowledges partial financial support of the DFG under grant no.\ Zi~209/6-1 and SFB~602. 

\input{pasref}

%
\end{document}

%% file: pasref.tex
%